\documentclass[a4paper]{article}

\usepackage{icrc2013}
\usepackage[english]{babel}
\title{Tau neutrino search with Cherenkov telescopes}

\shorttitle{Tau neutrino search with Cherenkov telescopes}

\authors{
Dariusz G\'ora$^{1,2}$,
Elisa Bernardini$^{1}$,
}

\afiliations{
$^1$ DESY, Platanenallee 6, D-15738 Zeuthen, Germany,\\
$^2$ Institute of Nuclear Physics PAN, Radzikowskiego 152, Cracow, Poland\\
}

\email{Dariusz.Gora@desy.de,Elisa.Bernardini@desy.de}

\abstract{Cherenkov telescopes could have the capability of detecting high
energy tau neutrinos by searching for very inclined
showers. If a tau lepton, produced by a tau neutrino,
escapes from the Earth crust, it will decay and initiate  an air shower which can be detected by a fluorescence/Cherenkov telescope.
Here we present a detailed Monte Carlo simulation of event rates induced by tau neutrinos in the energy range from 1 PeV to 1 EeV. Topographic conditions are taken into account for a set of example locations. As expected, we find a neutrino sensitivity which depends on the shape of the  energy spectrum from astrophysical sources.
We compare our findings with the sensitivity of the dedicated IceCube neutrino telescope under different conditions. We also find that a difference of several factors can be observed depending on the topographic conditions of the sites sampled.}

\keywords{Cherenkov telescopes, Tau Neutrinos.}

\begin{document}
\maketitle
\section{Introduction}
Many models which try to explain the origin of the ultra high energy cosmic rays (UHECR) 
claim  that their might be produced by Active Galactic Nuclei (AGN) and Gamma
Ray Bursts (GRB). Most of these models also predict a significant  flux of high energy neutrinos
from decay of charged pions. The chance of discovering extraterrestrial signals (neutrinos) from these objectes  largely varies with source classes and model predictions. For what concerns Blazars for example~\cite{Atoyan:2004pb}  Flat Spectrum Radio Quasars (FSRQ) are more promising, than BL-Lac objects, whereas in~\cite{PhysRevD.80.083008}  the opposite is predicted. The proton blazar model~\cite{Mucke2003593} predicts that the Low synchrotron peaked BL-Lacs (LBL) are more likely to produce a significant neutrino emission than the High synchrotron peaked BL-Lacs (HBL). On the other hand in \cite{Atoyan:2004pb} the considered p-$\gamma$ model leads to the conclusion that FSRQs bright in the GeV range 
are promising neutrino sources without any assumption on the spectral index.

From this point of view, detection of individual flares from AGNs, on the time scale of days 
or weeks can be more or less feasible for cubic-km scale neutrino telescopes like IceCube,
 based on different predictions for the mechanism yielding the observed electromagnetic emission at high energies.

 The highest energy neutrinos are expected to be born as muon and electron neutrinos, but due to  vacuum
oscillations a flux of high energy cosmic neutrinos at Earth is expected to be almost equally distributed among 
the three neutrino flavors.  Due to their low interaction probability, neutrinos
need to interact with a large amount of matter in order to be possibly
detected. The atmosphere and the Earth offer such a target. Since the
Earth is not transparent for neutrinos at the highest energies, one of
the detection techniques is based on the development of extensive air
showers (EAS) in the atmosphere.
In air, very inclined EAS can be detected only by instruments observing a
large volume. Propagating through the Earth only the so-called Earth
skimming tau neutrinos may initiate detectable air showers above the
ground. 
A successfull detection of such showers requires a ground array detector having 
a large acceptance, of the order of one km$^2$ and  a great sensitivity to horizontal showers 
 such as the  Pierre Auger Observatory, which is  sensitive to 
tau neutrinos in the  EeV energy  range~\cite{oscarprl::2008}.  
However, the detection of  PeV  tau neutrinos (expected to be produced by  AGNs and GRBs) 
through optical signals would also seem to be possible. A combination of fluorescence light and Cherenkov light detector in the shadow of steep cliff could achievie this goal ~\cite{doi:10.1142/S0217732304014458}.
Recently, it was also  shown, that such kind of experiments  could be  sensitive to tau neutrinos from fast transient
 objects like nearby GRBs~\cite{Asaoka:2012em}. 
\begin{figure*}[!t]
  \begin{center}                                                                                                    \includegraphics[width=\columnwidth,height=6.1cm]{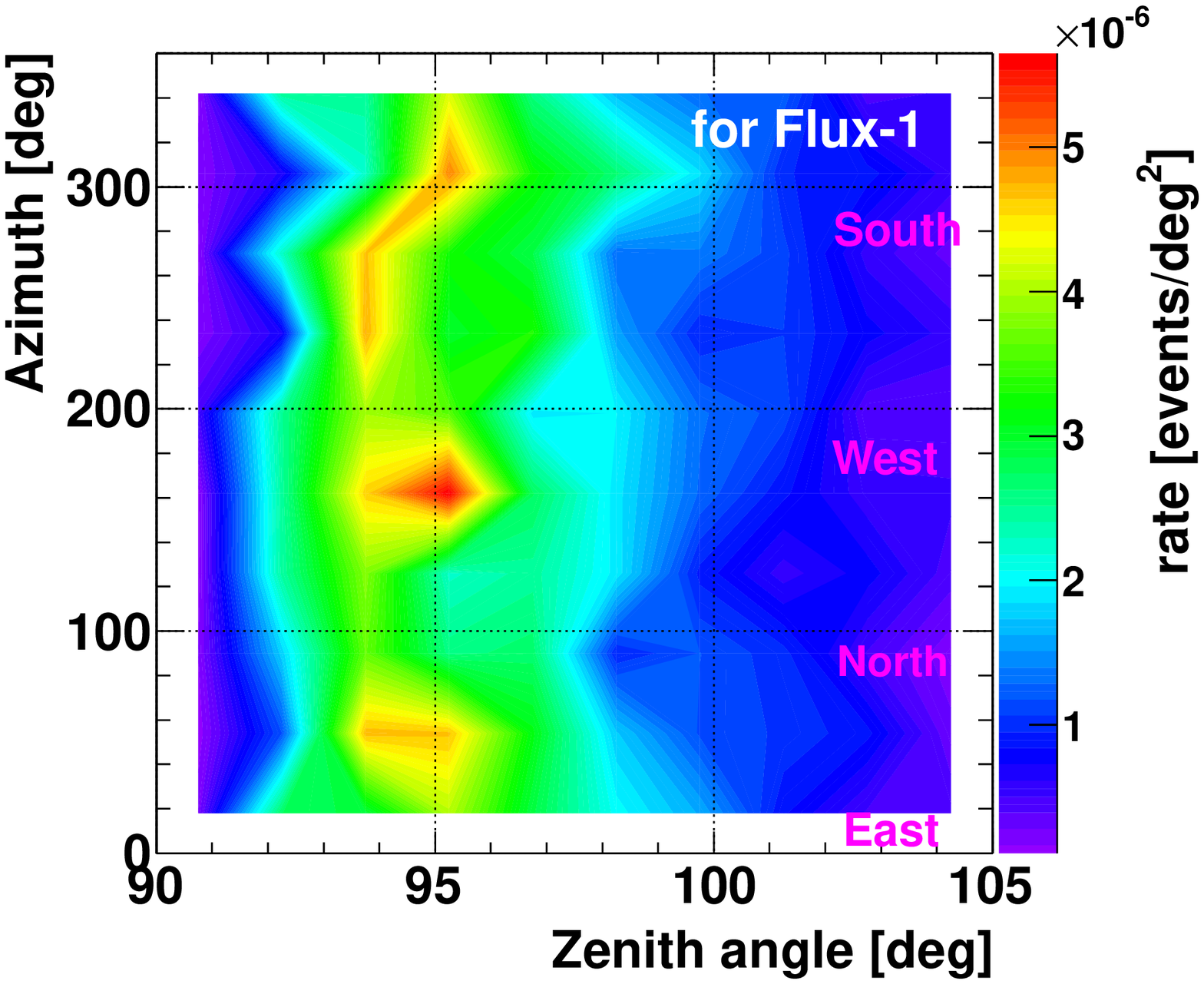} 
 \includegraphics[width=\columnwidth,height=6.1cm]{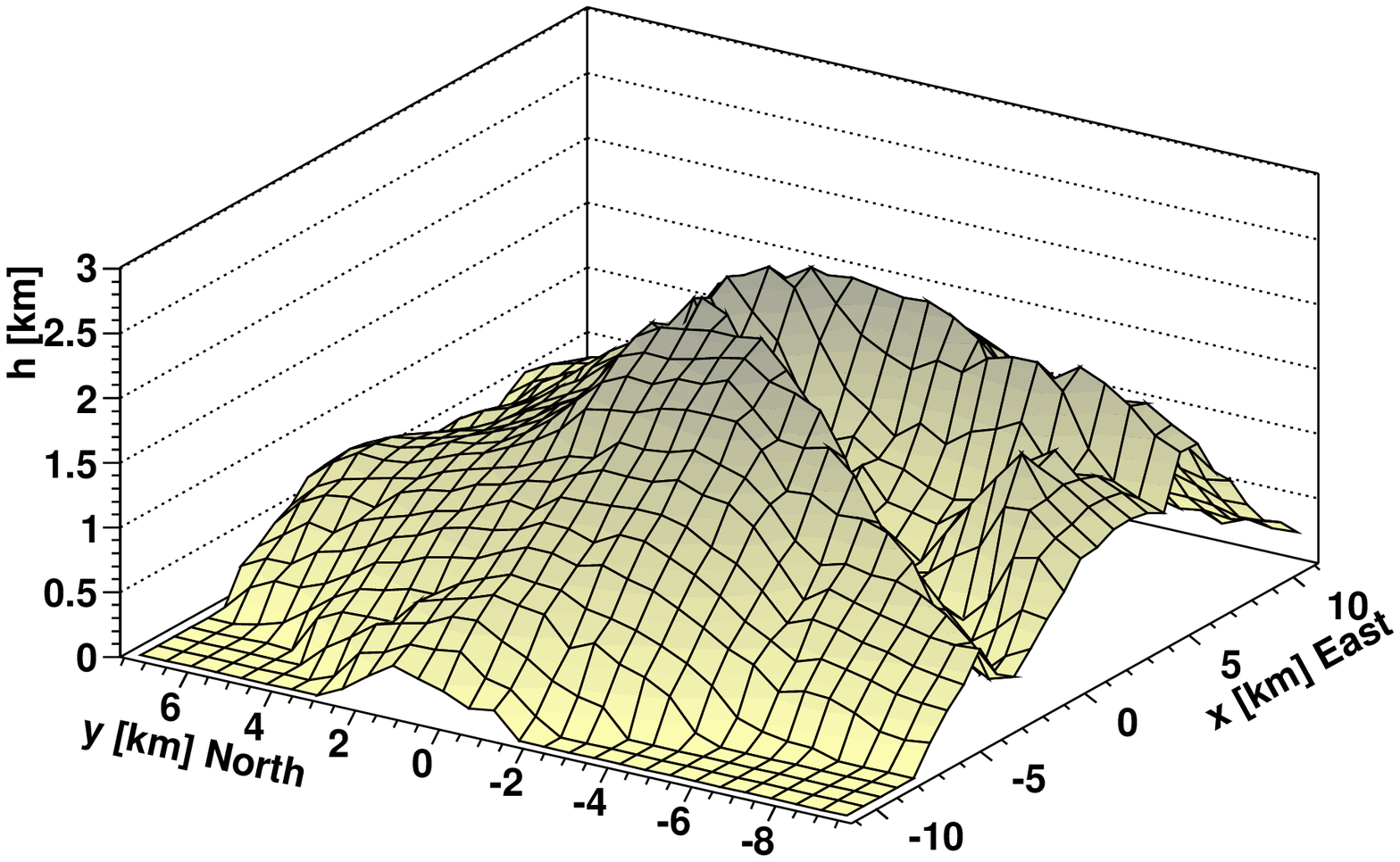}                                                       
  \end{center}                                                                                                               
  \caption{\label{fig::magic} {(Left panel) Expected  event rate as a function of azimuth and zenith in the case of
tau leptons with  energies between $10^{16}$ eV and $10^{17}$ eV;
(Right panel)  Topography of the La Palma Island according to CGIAR-CSI data. 
The center of the maps corresponds to the position of 
the MAGIC telescopes (latitude $\phi_{\mathrm{MAGIC}}=28^{\circ} 45' 34''$ N, longitude                 $\lambda_{\mathrm{MAGIC}}=17^{\circ} 53' 26''$ W, height 2200 m a.s.l.).}
 }                                                                                                                        
\end{figure*}   

In this work we investigate the detection of  high energy  tau neutrinos 
in the energy range from PeVs to EeVs by searching  for very inclined showers  using Cherenkov telescopes. 
We have  performed detailed  Monte Carlo (MC) simulations of expected tau neutrino event rates,
including local topographic  conditions, for  La Palma, i.e. the location of the  MAGIC telescopes~\cite{magic} 
and  sample selection of few  sites proposed  for the Cherenkov  Telescope Array (CTA)~\cite{cta}. 
Results  are shown  for  a few representative neutrino fluxes expected for giant flares from AGNs.
\begin{figure}[h]                                                                                                             
  \begin{center}                                                                                                              
    \includegraphics[width=\columnwidth]{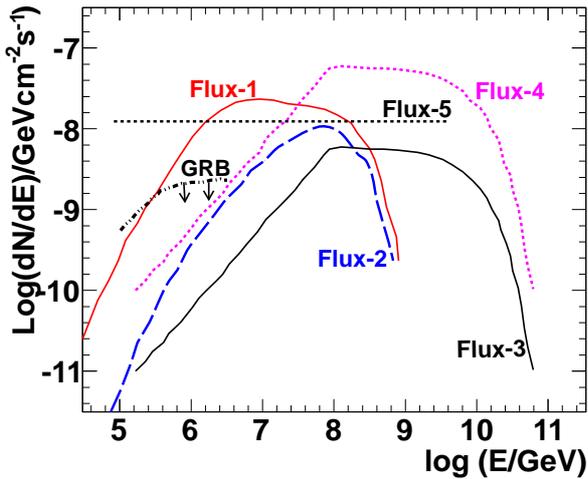}
  \end{center}                                                                                                               
  \caption{\label{fig::spectrum2} {A sample of representative neutrino fluxes from photo-hadronic interactions in AGNs. See text for more details.}} 
\end{figure}   

\section{Method}

The propagation of a given neutrino flux
 through the Earth and the atmosphere  is simulated using  an extended version of the
code ANIS~\cite{gora:2007}.

For fixed neutrino energies, $10^{6}$ events are generated on  top of the atmosphere with zenith
angles ($\theta$) in the range $90^{\circ}$--$105^{\circ}$ (up-going showers) and with azimuth angles
in the range $0^{\circ}$--$360^{\circ}$. Neutrinos are propagated along their trajectories of
length $ \Delta L$ from the generation point on  top of the
atmosphere to the backside of the detector in steps of $\Delta L$/1000
($\geq 6$ km). At each step of propagation, the $\nu$--nucleon
interaction probability is calculated according to different
parametrization of its cross section based on the chosen parton
distribution function (PDF).
In particular, the propagation of tau leptons through the Earth is
simulated with different energy loss models. 
All the computations are done using digital elevation maps
(DEM)~\cite{dem} to model the surrounding mass distribution of each considered site. 
The flux of the outcoming leptons as well as their
energy and the decay vertex positions are calculated inside a defined
detector volume (set to $35\times10$ km$^{3}$).
The acceptance for a given initial neutrino energy
$E_{\nu_\tau}$ is given by:
\begin{eqnarray}
  A(E_{\nu_\tau})  =N_{\mathrm{gen}}^{-1} \times \sum_{i=1}^{N_{k}}
   P_{i}(E_{\nu_\tau},E_{\tau},\theta) \nonumber \\ 
   \times T_{\mathrm{eff},i}(E_{\tau},x,y,h,\theta) \times
   A_i(\theta)\times \Delta \Omega,
\label{aperture}
\end{eqnarray}
where $N_{\mathrm{gen}}$ is the number of generated neutrino events
and $N_k$ is the number of $\tau$ leptons with energy $E_{\tau}$ larger then the threshold $E_{\mathrm{th}}>1$ PeV and decay vertex position
inside the detector volume. $P(E_{\nu_\tau},E_{\tau},\theta)$ 
is the probability that a neutrino with
energy $E_{\nu_\tau}$ and crossing the distance $\Delta l$ would produce a
particle with an energy $E_{\tau}$ (this probability was used as
"weight" of the event), $A_i(\theta)$ is the cross-sectional area of the detector volume seen by the neutrino,
 $\Delta\Omega$ is the space angle.  The $T_{\mathrm{eff}}(E_{\tau},x,y,h,\theta)$ is 
the trigger efficiency   for   tau lepton induced showers with 
first interaction position (x, y) and height (h) above the ground. 
The  trigger efficiency depends on the response  of a given detector, and is usually 
 estimated based on MC simulations. In this work we used  an average trigger efficiency  
extracted from~\cite{Asaoka:2012em}, namely $\langle T_{\mathrm{eff}} \rangle =10$\%, which is   comparable  with  what 
calculated for  up-going tau neutrino showers studied in~\cite{doi:10.1142/S0217732304014458}.
This is a qualitative estimation  and as such it is the major source of uncertanties 
on the results presented hereafter.

 Eq.~(\ref{aperture})  gives the  acceptance for diffuse  neutrinos. 
The acceptance for a given point source could be estimated as the ratio between the diffuse acceptance  
and the solid angle covered by the diffuse analysis, multiplied by the fraction of time the source is visible $f_{\mathrm{vis}}(\delta_{s},\phi_{\mathrm{site}})$ with the aperture defined in the beginning.
This fraction depends on source declination ($\delta_{s}$) and the latitude of the observing site ($\phi$). 
In this work the point source acceptance is  calculated as: 
 $ A^{\mathrm{PS}}(E_{\nu_\tau})\simeq \frac{A(E_{\nu_\tau})}{\Delta \Omega}\times f_{\mathrm{vis}}(\delta_{s},\phi_{\mathrm{site}})$.
\begin{table*}[bt!]
  \caption{\label{tab::rate222} {Expected event rate, $N_\mathrm{La Palma}$  for
     a detector located at the La Palma Roque with a trigger efficiency of 10\%, compared with  what expected for 
IceCube (with realistic efficiency). The values are calculated with ALLM~\cite{allm} tau energy loss model
and GRV98lo~\cite{GRVlo} cross-section, with  $f_{\mathrm{vis}}=100$\%, $\Delta \Omega=2\pi (\cos(90^{\circ})-\cos(105^{\circ}))=1.6262$ and  $\Delta T=3$ hours. The rate are calculated with 
the point source acceptances shown in Figure~\ref{fig::acccc}.}}
\begin{center}
\begin{tabular}{ccccccccc}
\hline
\hline
 &Flux-1  &Flux-2&  Flux-3 & Flux-4 &Flux-5 &GRB\\
\hline
\hline
$N_{\mathrm{La Palma}}$  &0.00028 & 0.00015 & $8.6 \times 10^{-5}$&0.00086 &0.00026& $0.50\times 10^{-5}$\\ 
&    &   &      &  & &\\
$N^{\mathrm{Northern \mbox{ } Sky}}_{\mathrm{IceCube}}$&  0.00068&  0.00025 & $4.6\times 10^{-5}$  & 0.00046 & 0.00088 &$4.4\times10^{-5}$\\
$N^{\mathrm{Southern \mbox{ } Sky}}_{\mathrm{IceCube}}$&  0.00110&  0.00032 & $7.6\times 10^{-5}$  & 0.00076 & 0.00088 &$3.4\times10^{-5}$\\
\hline
\hline
\end{tabular}
\end{center}
\end{table*}
In Figure~\ref{fig::spectrum2}  a compilation of fluxes expected from AGN flares are shown.
 Flux-1 and Flux-2 are  calculations  for the February 23, 2006 $\gamma$-ray flare of 3C 279~\cite{2009IJMPD}.
Flux-3  and Flux 4  predictions for PKS 2155-304  
in low-state and high-state, respectively~\cite{Becker2011269}. 
Flux-5 corresponds to a prediction  for 3C 279 calculated in~\cite{PhysRevLett.87.221102}. The  flux labeled as 
GRB corresponds to the recent limit on the neutrino emission 
from GRBs reported by the IceCube Collaboration~\cite{Abbasi:2012zw}.

The total observable
rates (number of expected events) were calculated as $N=\Delta T \times \int_{E_{\mathrm{th}}}^{E_{\mathrm{max}}}
A^{\mathrm{PS}}(E_{\nu_\tau})\times\Phi(E_{\nu_\tau})\times dE_{\nu_\tau}$, where $\Phi(E_{\nu_\tau})$ is the
neutrino flux and $\Delta T$  an arbitrary observation time (3 hours in Table~\ref{tab::rate222}).
\section{Results} \label{sec:results}
\begin{figure}[h]                                                                                                                    
  \begin{center}                                                                                     
    \includegraphics[width=\columnwidth]{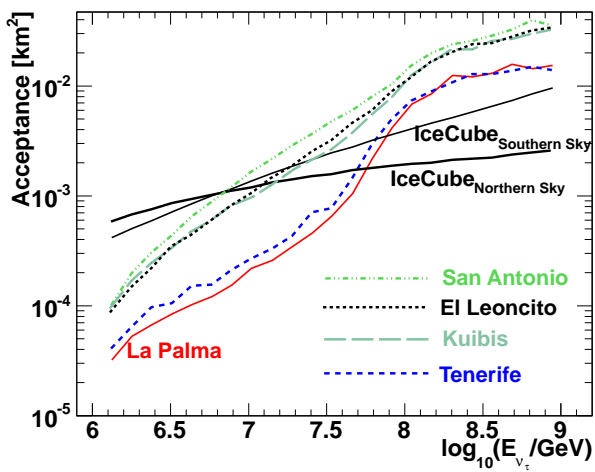}
  \end{center}
\vspace{-0.5cm}
  \caption{\label{fig::acccc} Acceptance, $A^{\mathrm{PS}}(E_{\nu_\tau})$ to  earth-skimming tau  neutrinos  as estimated  for the  La Palma
  site  and  a sample selection of few CTA sites (with a trigger efficiency of 10\%) and  IceCube (with  correct efficiency, as extracted from~\cite{IC-80-acc}). } 
\end{figure}

Figure~\ref{fig::magic} shows the expected rate for  a detector located at the La Palma site
 (with an average trigger efficiency of 10\%) as a function of the incoming neutrino direction 
(defined by the zenith and azimuth angles). Correlations can be observed
 between the expected  rate  and the local topography i.e. the expected number of events 
from South-East is usually larger than from other directions,
 due to largest amount of matter encountered by incoming neutrinos when coming from the South-East direction.
For  tau lepton energies between $10^{16}$ eV  and $10^{17}$ eV the decay length   
is  a few kilometers, so detectable events   should mainly  come from local  hills 
from La Palma Island when considering the location of MAGIC.
While for tau leptons  energies between $10^{18}$ eV  and $10^{19}$ eV the decay length  is
larger than 50 km, so that the matter distribution of other Canary Islands  can also slightly contribute.

In Figure~\ref{fig::acccc} we show the estimated point source acceptance  for the La Palma  site 
and other possible locations as a function of the neutrino energy together with the  IceCube acceptance as extracted from~\cite{IC-80-acc}. We stress at this point that we aim at exploring the effect of different 
topographic conditions rather than providing a comprehensive survey of potential sites.

The IceCube acceptance  shows an increase 
for energies between $10^{6}$ GeV and $10^{9}$ GeV, and  is on  average  
about $~2\times10^{-3}$ km$^{2}$. A potential detector located in La Palma
with an average trigger efficiency of 10\% can have an acceptance as large as 
a factor 5 greater than IceCube (Northern Sky) at energies larger than  $\sim 5\times 10^{7}$ GeV. 
Indeed,  for  neutrino fluxes  covering the energy range  below $\sim 5\times10^{7}$ GeV 
(Flux-1, Flux-2, Flux-5 and GRB) the number of expected  events is smaller 
than  what estimated  for IceCube  assuming 3 hours of observation time. 
However, even in this energy range (see~Table~\ref{tab::rate222})  similar rates  can be 
obtained with a trigerr efficiency increased by a  factor 2-3 compared to the rough estimate of 10\%.
For  Flux-3 and Flux-4 the event rate is  about a factor 2 larger than the realistic
rate calculated for IceCube (Northern Sky). This indicates  that Cherenkov telescopes
 could have a sensitivity comparable or even  larger  to 
 neutrino telescopes such as IceCube in case of  short  neutrino  flares 
(i.e.  with a duration of about  a few hours). For larger durations the advantage of neutrino telescopes
as full sky no  dead-time  instruments will be relevant. An accurate simulation of the neutrino trigger efficiency 
for realistic  Cherenkov telescopes is however needed. Table~\ref{tab::rate222} 
 also shows that for a  GRB flux at the level of the current IceCube 
limit a trigger efficiency larger by at least a factor 10 compared to what we assumed here is needed.

\begin{figure*}[]
  \begin{center}
  \includegraphics[width=0.68\columnwidth,height=4.5cm]{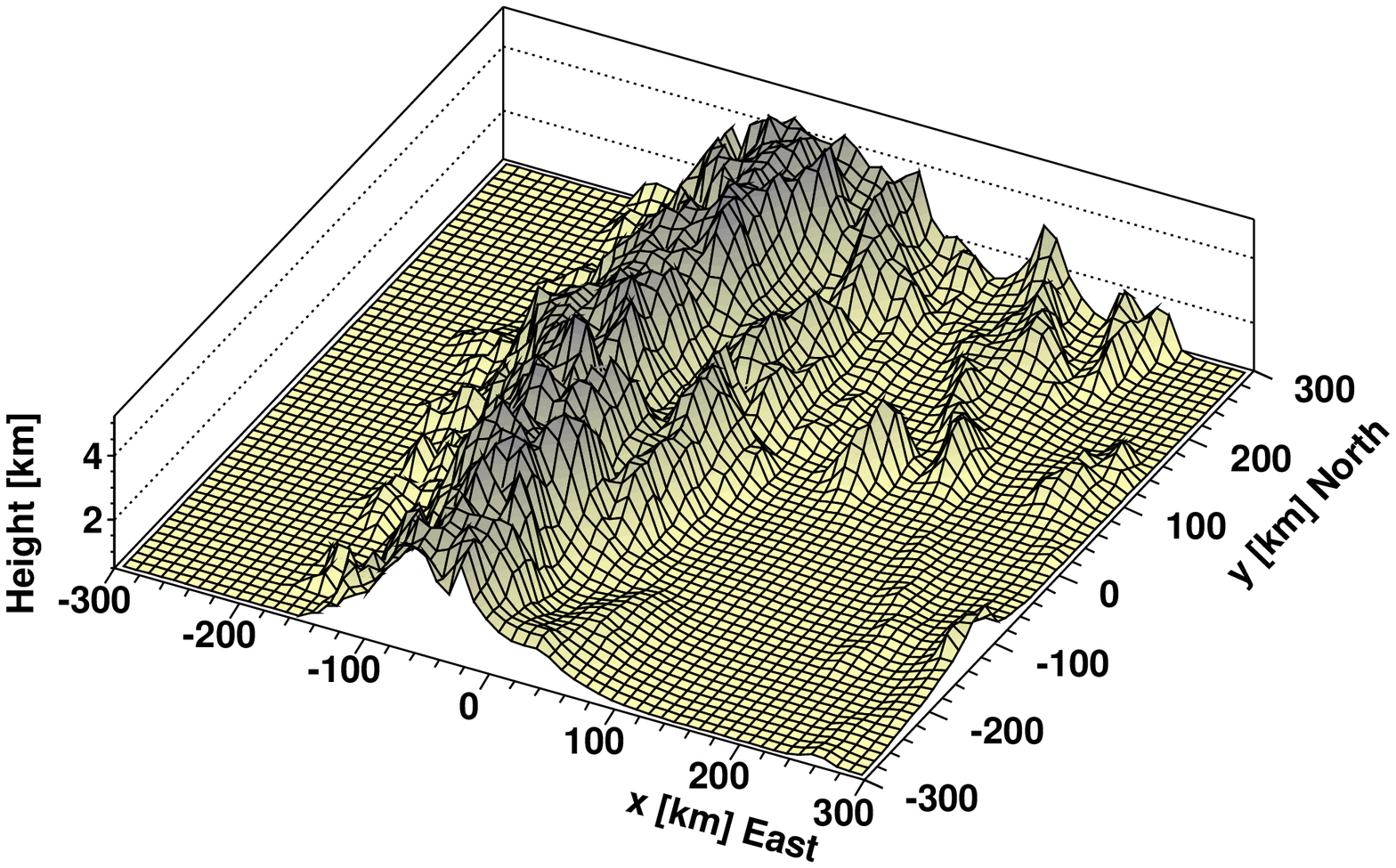}                 
  \includegraphics[width=0.68\columnwidth,height=4.5cm]{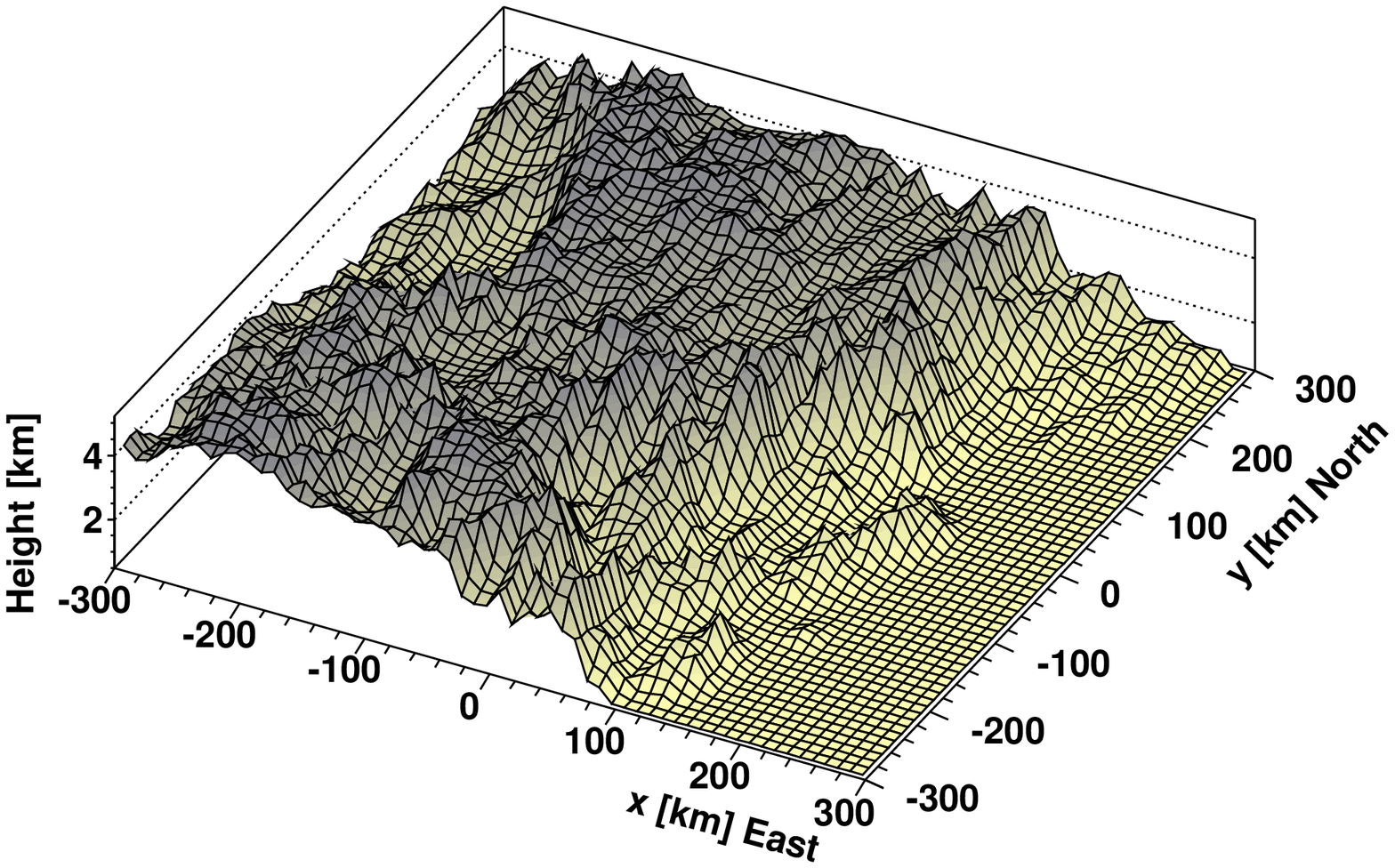}     
  \includegraphics[width=0.68\columnwidth,height=4.5cm]{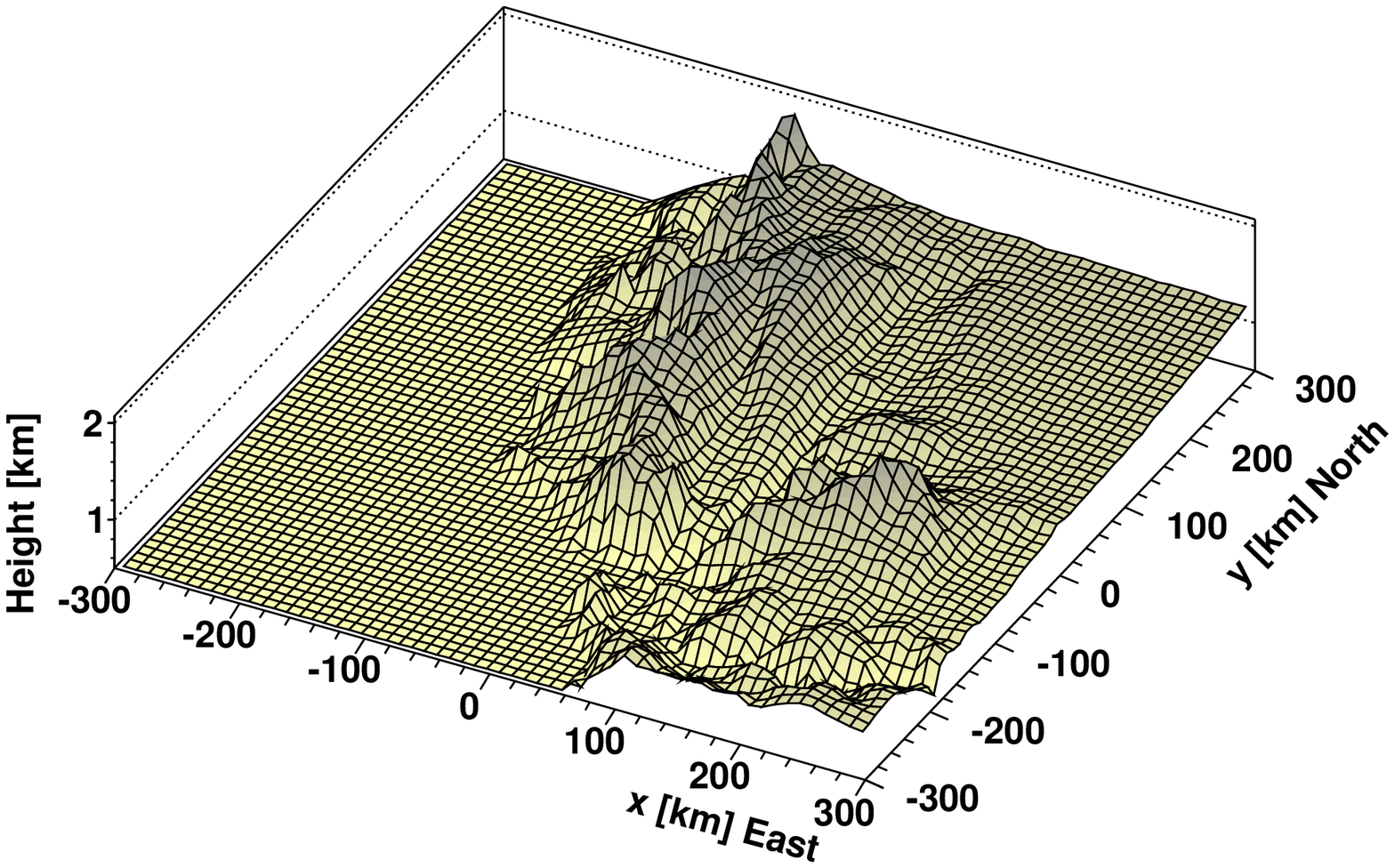}                 
  \end{center}                                                                                                              
  \caption{\label{fig::cta} Topography of the CTA site according to                                             
    CGIAR-CSI data. The center of the map corresponds to the center of                                           
    the site; (left) El Leoncito, Argentina (latitude $\phi_{\mathrm{Center}}=31^{\circ} 41' 18.6''$ N, 
 longitude $\lambda_{\mathrm{Center}}=69^{\circ} 16' 58.2''$ W); (middle) San Antonio, Argentina (latitude $\phi_{\mathrm{Center}}=24^{\circ} 02' 42.7''$ N, longitude                                                          $\lambda_{\mathrm{Center}}=66^{\circ} 14' 5.8''$ W); (right) Kuibis, Namibia (latitude $\phi_{\mathrm{Center}}=26.68333^{\circ}$ N, longitude                                                        $\lambda_{\mathrm{Center}}=16.88333^{\circ}$ E).   
}                                                                                                                        
\end{figure*}   
An other interesting possibility, for the detection  of up-going tau neutrinos,
is to built Cherenkov detectors at sites surrounded by mountains. Mountains can work as  additional target 
and  will  lead to an enhancement of emerging tau leptons.
A target  mountain can also function as a shield to  cosmic rays and star light.

In order to estimate the possible influence of mountains on the calculated event rate for up-going tau neutrinos,
 we performed  a similar simulation  as done for La Palma  site  for four  sample locations:
 two in the Argentina (San Antonio, El Leoncito), one in Namibia (Kuibis) and one in the Canary Islands
(Tenerife), see  Table~\ref{tab::rate_cta} and Figure~\ref{fig::cta}. In case of sites surrounded 
by mountains (San Antonio, El Leoncito, Kuibis)  results show an higher event rate
(by at least a factor of 2) than for a site  without   surrounding mountains (La Palma and Tenerife).
\begin{table}[h]
  \caption{\label{tab::rate_cta} {The ratio of event rates defined as: $k = N^{i}_{\mathrm{site}}/ N_{\mathrm{La Palma}}$  for the Flux-1 and Flux-3.}}
\vspace{0.2cm}
\begin{center}
\small
\begin{tabular}{cccc}
\hline
\hline
  &       Site & $k_{\mathrm{Flux-1}}$  &$k_{\mathrm{Flux-3}}$  \\
\hline
 & \bf San Antonio  & 4.6 &2.6   \\
 & \bf El Leoncito  & 3.2 &2.1   \\
 & \bf Kuibis       & 2.7 &1.9  \\
 & \bf Tenerife     & 1.2 &1.1  \\
\hline
\hline
\end{tabular}
\end{center}
\end{table}

We also studied the influence on the expected event rate arising from 
uncertainties on the tau lepton energy  loss.
The average energy loss of taus per distance travelled 
(unit depth $X$ in  gcm$^2$) can be described as
$ \left\langle \frac{dE}{dX} \right \rangle = \alpha(E) + \beta(E) E $.
The factor $\alpha(E)$, which is nearly constant, is due to ionization
 $\beta(E)$ is the sum of $e^+e^-$-pair production and bremsstrahlung,
  which are both well understood, and photonuclear scattering,
  which is not only the dominant contribution at high energies but at the same time subject to relatively large uncertainties. In this work the  factor $\beta_{\tau}$ are  calculated using 
the  following models describing contribution of photonuclear scattering:
ALLM~\cite{allm}, BB/BS~\cite{bbbs} and CMKT~\cite{ckmt}. 
and  different neutrino-nucleon cross-sections:  GRV98lo~\cite{GRVlo}, 
CTEQ66c~\cite{cteq}, HP~\cite{hp}, ASSS~\cite{CooperSarkar:2007cv}, ASW~\cite{Albacete:2005ef}. 
Results are listed in~Table~\ref{tab::rate_cta} for  Flux-1 and Flux-3.

\begin{table}[h]
 \caption{\label{tab::rate} {Relative contributions to the systematic uncertainties on
the up-going tau neutrino rate. As a reference  value the expected event rate
for La Palma site  calculated for Flux-1 and Flux-3 (in brackets) was used.}}
\small
\center 
\begin{tabular}{ccccc}
\hline
\hline
rate &  PDF  & $\beta_{\tau}$ & sum     \\
\hline
\hline
             &+14\% (+42\%)   &+2\% (+7\%) &  +14\% (+43\%)   \\
$2.8\times 10^{-4}$  &             &             &      \\
  ($8.6\times 10^{-5}$)           &-2\% (-7\%) &-7\% (-14\%)  & -7\% (-16\%)  \\
\hline
\hline
\end{tabular}
\end{table}
\section{Summary}
In this paper detailed Monte Carlo simulation of event rate, including local topographic  conditions of the detector, and  using  recent predictions for neutrino fluxes in AGN flares
are presented for La Palma site
and a few proposed CTA  sites. The calculated  neutrino rate 
 is usually  worse compared to what estimated  for  IceCube 
assuming realistic observation  times spend by Cherenkov telescopes (a few hours). However  for models 
which predict neutrino fluxes with energy above  $\sim  5\times 10^{17}$ eV,  the sensitivity
 can be comparable to IceCube or even better.  For the sites considered which have surrounding  mountains the 
 expected event rate  is up to factor  5  higher compared to what expected for La Palma.


\end{document}